%% file: db.tex
\documentclass[submission,copyright,creativecommons]{eptcs}
\providecommand{\event}{FICS 2015} 

\usepackage{amsmath}
\usepackage{amssymb}
\usepackage{txfonts}
\usepackage{meq}
\usepackage{xspace}
\usepackage{theorem}

\usepackage[UglyObsolete]{diagrams}
\usepackage{enumitem}
\usepackage{wrapfloat}
\usepackage{fancyvrb}
\usepackage[dvipdfm]{graphicx}

\usepackage{header}

\begin{document}
\pagestyle{plain}

\title{Iteration Algebras for UnQL Graphs and\\
Completeness for Bisimulation}

\author{Makoto Hamana
\institute{Department of Computer Science, Gunma University, Japan}
  \email{hamana@cs.gunma-u.ac.jp}
}
\def\titlerunning{Iteration Algebras for UnQL Graphs}
\def\authorrunning{M. Hamana}
\maketitle

\input{abs}
\input{intro}

\input{theory}

\input{pr}
\input{alg}
\input{bex}
\input{concl}
\input{ack}

\bibliographystyle{eptcs}
\bibliography{bib}

\end{document}

%% file: abs.tex
\begin{abstract} 

This paper shows an application of Bloom and \Esik's iteration algebras to
model graph data in a graph database query language.  About twenty years
ago, Buneman et al. developed a graph database query language UnQL on the top
of a functional meta-language UnCAL for describing and manipulating
graphs. Recently, the functional programming community has shown renewed
interest in UnCAL, because it provides an efficient graph transformation
language which is useful for various applications, such as bidirectional
computation.  However, no mathematical semantics of UnQL/UnCAL graphs has
been developed. In this paper, we give an equational axiomatisation and
algebraic semantics of UnCAL graphs. The main result of this paper is to
prove that completeness of our equational axioms for UnCAL for the original
bisimulation of UnCAL graphs via iteration algebras.  Another benefit of
algebraic semantics is a clean characterisation of structural recursion on
graphs using free iteration algebra.

\end{abstract}


%% file: intro.tex
\section{Introduction}\label{sec:review}

Graph database is used as a back-end of various web and net services,
and therefore it is one of the important software systems in the Internet
society. About twenty years ago, Buneman et al. \cite{SIGMOD1996,Buneman2,Buneman} developed a graph
database query language UnQL (Unstructured data Query Language)
on top of a functional meta-language
\textbf{UnCAL} (Unstructured Calculus)
for describing and manipulating graph data. 
The term ``unstructured'' is used to refer to
unstructured or semi-structured data, i.e.,
data having no assumed format 
in a database
(in contrast to relational database).
Recently, 
the functional programming community 
found a new application area of UnCAL
in so-called bidirectional transformations on graph data,
because it provides an efficient graph transformation
language.
The theory and practice of
UnCAL 
have been extended and refined in various directions (e.g. \cite{ICFP10,LOPSTR,ICFP13,PPDP13}), which
has increased the importance of UnCAL.

In this paper, 
we give a more conceptual understanding of UnCAL using 
semantics of type theory and fixed points.
We give an equational axiomatisation  and algebraic semantics
of UnCAL graphs.
The main result of this paper is to prove completeness of our
equational axioms for UnCAL for the original bisimulation of UnCAL graphs via
iteration algebras.
Another benefit of algebraic semantics is 
a clean characterisation of 
the computation mechanism of UnCAL called
``structural recursion on graphs'' 
using free iteration algebra.

\input{ireview}


%% file: ireview.tex
\begin{rulefigw}
  \begin{center}
{\includegraphics[scale=.53]{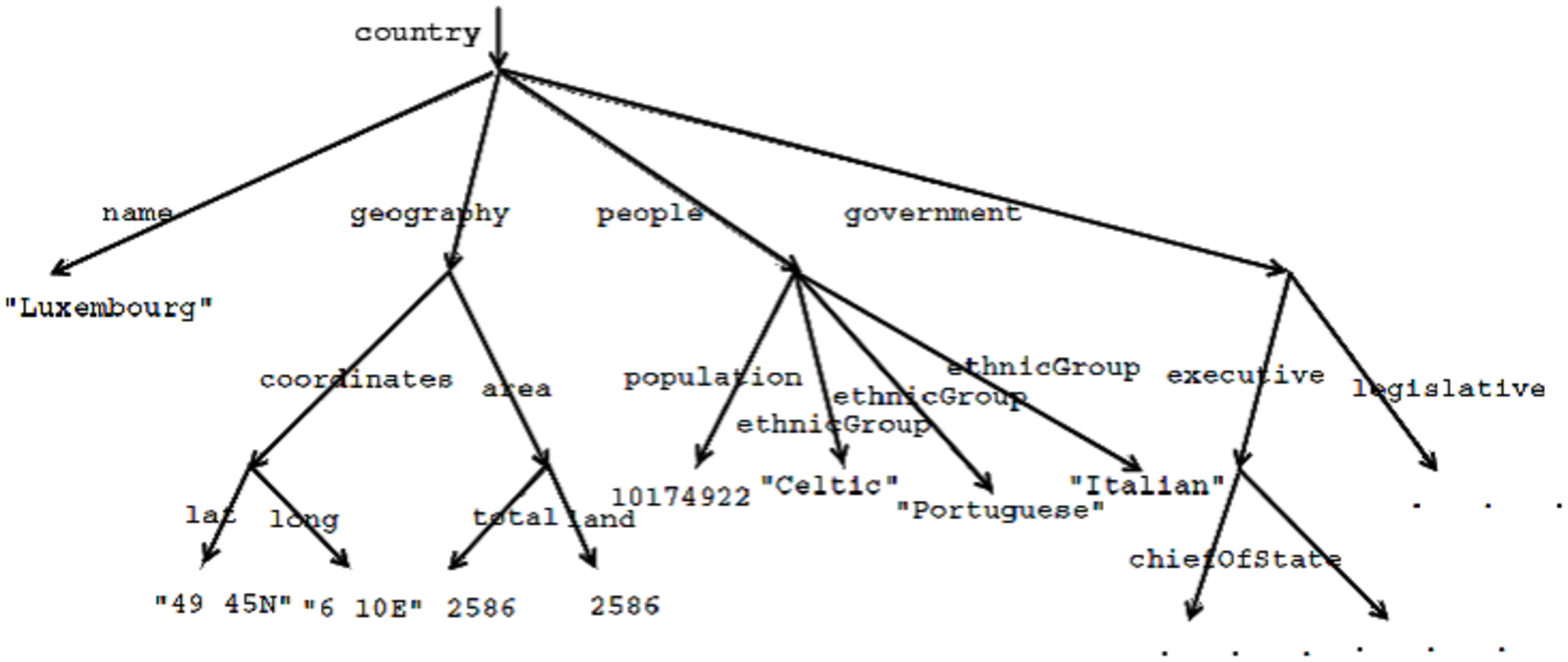}}      
  \end{center}
\y{-1em}
\end{rulefigw}

\subsec{UnCAL Overview}
We begin by introducing UnCAL.
UnCAL deals with graphs in a graph database. 
Hence, it is better to start
with viewing how concrete semi-structured data
is processed in UnCAL. Consider the
semi-structured data \code{sd} below
which is 
taken from \cite{Buneman}.
\newpage
It contains information about country, e.g. geography, people, 
government,
etc. 
\begin{wrapfigure}[8]{r}{.68\linewidth}\y{-1em}
\begin{Verbatim}[fontsize=\small,frame=single,baselinestretch=.9,
commandchars=\\\[\],codes=\mathcom]
sd $\;\triangleleft\;\,$ country:{name:"Luxembourg",
 geography:{coordinates:{long:"49 45N", lat:"6 10E"},
            area:{total:2586, land:2586}},
 people:{population:425017,
         ethnicGroup:"Celtic",
         ethnicGroup:"Portuguese",
         ethnicGroup:"Italian"},
 government:{executive:{chiefOfState:{name:"Jean",..}}}}
\end{Verbatim}
\end{wrapfigure}
\noindent
It is depicted as a tree above,
in which  edges and leaves are labelled.
Using UnCAL's term language for describing graphs (and trees),
this is 
defined by \code{sd} shown at right.
Then we can define functions in UnCAL to 
process data. For example, a function that retrieves all ethnic groups in the graph can be 
defined simply by 
\begin{Verbatim}
      sfun f1(L:T) = if L = ethnicGroup then (result:T) else f1(T)
\end{Verbatim}

\noindent
The keyword \code{sfun} denotes a function definition by 
\Hi{structural recursion on graphs}, which is
the computational mechanism of UnCAL. 
Executing it, we can certainly extract:  
\begin{Verbatim}[commandchars=\\\[\],codes=\mathcom]
  f1(sd) \narone {result:"Celtic", result:"Portuguese", result:"Italian"}
\end{Verbatim}
\begin{wrapfigure}[16]{r}{.6\linewidth}\y{-.5em}
\x{-1.5em}\includegraphics[scale=.44]{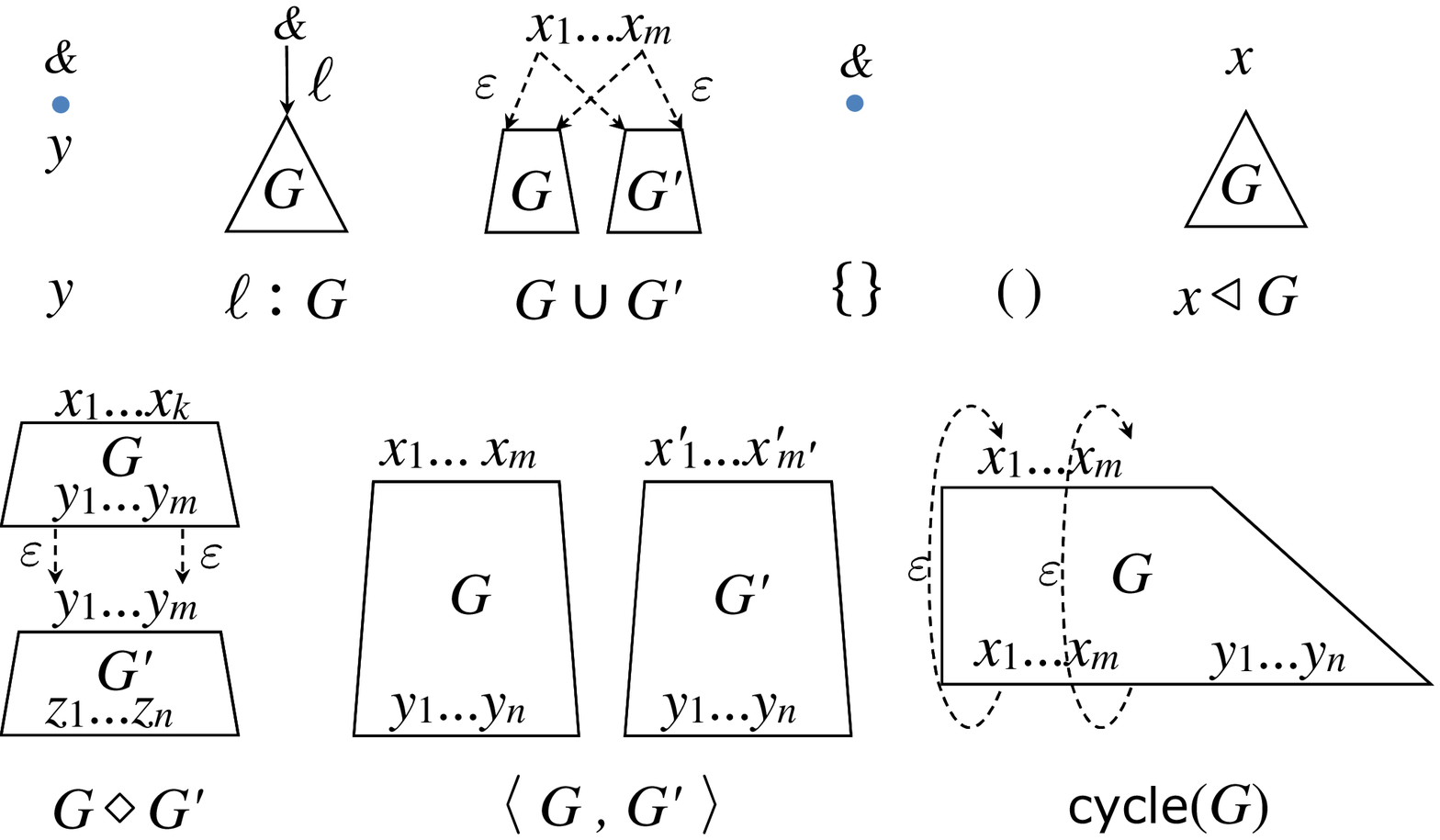}  
\caption{\small Graph theoretic definitions of constructors {\scriptsize
    \cite{Buneman}\\ 
Slightly changed notation. Correspondence
between the original and this paper's:\\
$\&y = y,\quad @ = \at,\quad \oplus=<-,->,\quad (-:=-) = \nxsub{-}{-}$.
}}
\label{fig:graph-constr}
\end{wrapfigure}

The notation {\small \verb!{!$\ccc$\code{:}$\ccc,\ccc$\verb!}!} 
is a part of the UnCAL's term language for representing graphs.
It consists of markers $\var x$, 
labelled edges ${\,\ell\co{t}\,}$,
vertical compositions  $s \at t$,
horizontal compositions   $\pa{s\opl t}$,   
other horizontal compositions 
${s} {\;\RM{\scriptsize $\union$}\;} {t}$ merging roots,
forming cycles $\cy{t}$,
constants $\nil$,$\EMP$, and definitions $\xsub{x}{t}$.
These term constructions have underlying graph 
theoretic meaning shown at th right.
Namely, these are officially defined as operations 
on the ordinary representations of graphs:
(vertices set, edges set, leaves, roots)-tuples 
$(V,E,\set{\var{y_1},\ooo\var{y_m}},\set{\var{x_1},\ooo,\var{x_n}})$,
but we do not use 
the graph theoretic definitions of
these operations in this paper. 

\newpage
UnCAL deals with graphs 
\Hi{modulo bisimulation} (i.e. not only modulo graph isomorphism).
\begin{wrapfigure}[14]{l}{.35\linewidth}
\x{-1em}\includegraphics[scale=.38]{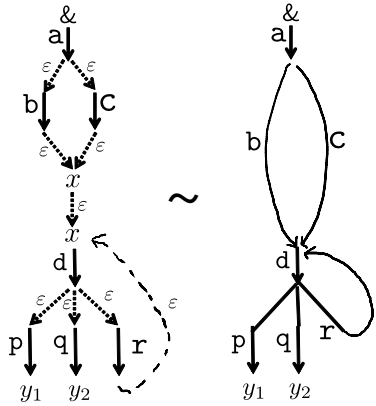}  
\caption{\small Graph $G$ and bisimilar one}
\label{fig:G}
\end{wrapfigure}
An UnCAL graph is directed and have (possibly multiple) root(s)
written $\dmark$ (or multiple $\var{x_1}\ccc\var{x_n}$) and leaves (written $\var{y_1}\ccc\var{y_m}$),
and with
the roots and leaves
drawn pictorially at the top and bottom, respectively.
The symbols $x,y_1,y_2,\dmark$ in the figures and terms
are called markers, which are
the names of nodes in a graph
and are used for references for cycles.
Also, they are used as port names to connect two graphs.
A dotted line labelled \epsilon
is called an \epsilon-edge, which is a ``virtual'' edge 
connecting two nodes 
directly.
This is achieved by identifying graphs by \Hi{extended bisimulation},
which ignores \epsilon-edges suitably in UnCAL.
The UnCAL graph $G$ 
shown at the left
is an example. 
This is extended bisimilar to a graph that reduces all \epsilon-edges.
Using UnCAL's language, 
$G$ is represented as the following term $\mathbf{t}_G$ 
$$
\mathbf{t}_G \; =\; 
 \code a \cc{\Uni{\code b\co{\var x}}{\code c\co{\var x}}} \;\at\;
 {\cy{\nxsub{x}{\code  d\cc{ \UniT{\code p\co{\var{y_1}}} {\code q\co{\var{y_2}}} {\code r\co{\var{x}}}}
   \;} }}.$$
UnCAL's structural recursive function works also on cycle.
For example, define another function 
\begin{Verbatim}
                      sfun f2(L:T) = a:f2(T)
\end{Verbatim}

\noindent
that
replaces every edge with \code a. 
As expected,
$$\code{f2}(\,\mathbf{t}_G\,) \;\;\narone\;\; \code a \cc{\Uni{\code a\co{\var x}}{\code a\co{\var x}}} \;\at\;
   {\cy{\nxsub{x}{ \code a\cc{ \UniT{\code a\co{\var{y_1}}} {\code
   a\co{\var{y_2}}} {\code a\co{\var{x}}}}
   \;} }}$$
where all labels are changed to \code{a}.

Another characteristic role of bisimulation is 
that it identifies expansion
of cycles.
For example, 
a term $\cy{\nxsub \dmark{{\code a}\co \dmark}} $ corresponds to
the graph shown below at the leftmost.
It is bisimilar to the right ones, especially
the infinitely expanded graph shown at the rightmost, which has no cycle.
\medskip
\begin{center}
\ \qquad\includegraphics[scale=.37]{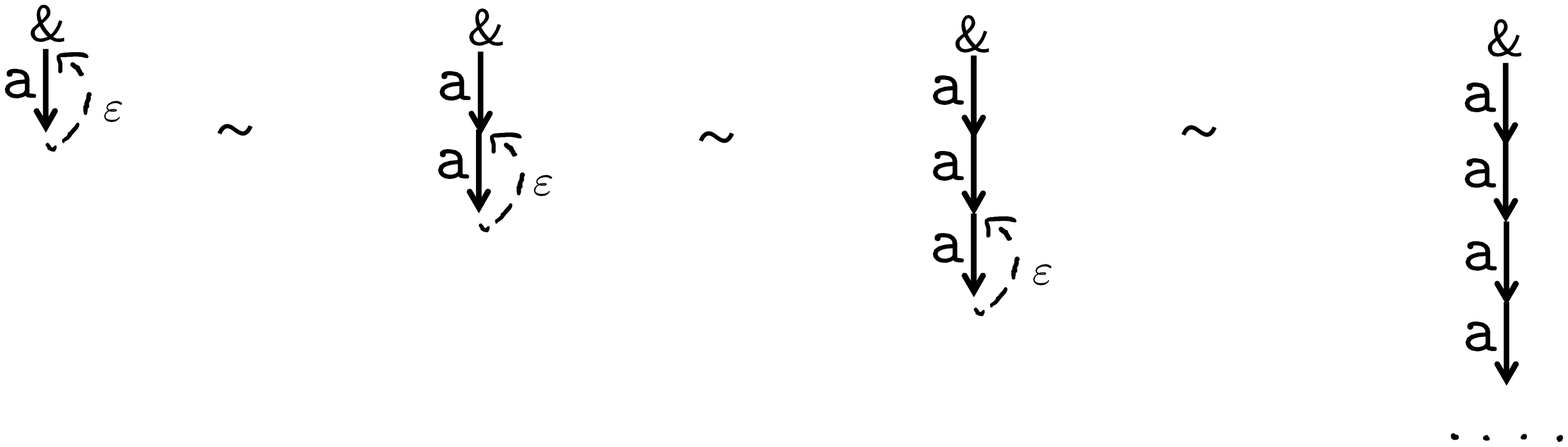}  
\end{center}
\y{-2em}
These are in term notation:
\y{-.5em}
$$\cy{\nxsub \dmark{{\code a}\co \dmark}}
\;\;\bisim\;\;  \code a \co \cy{\nxsub \dmark{{\code a}\co \dmark}}
\;\bisim\;  \code a \co \code a \co \cy{\nxsub \dmark{{\code a}\co \dmark}}
\ \x{5em}
$$

\subsec{Problems}
There have been no algebraic laws that establish the above expansion of
\cyc. Namely, these are merely bisimilar, and not a consequence of 
any algebraic law.
But obviously,
we expect that
it should be a consequence of the algebraic law of \Hi{fixed point property} of \cyc.

In the original and subsequent formulation of UnCAL 
\cite{Buneman,ICFP13,ICFP10,PPDP13},
there are complications of this kind.
The relationship between terms and graphs in UnCAL
is not a one-to-one correspondence.
No term notation exits for \epsilon-edges and 
infinite graphs (generated by the cycle construct), thus
the rightmost infinite graphs of the above expansion cannot be expressed in syntax.
But such an infinite graph is allowed as a possible graph in 
the original formulation of UnCAL.
Consequently, instead of terms,
one must use graphs and graph theoretic reasoning with
care of bisimulation to reason about UnCAL.
Therefore, a property in UnCAL could not be established 
only using induction on terms.
That fact sometime makes some proofs about UnCAL
quite complicated.

Because UnCAL graphs are identified by bisimulation,
it is necessary to
use a procedure or algorithm to check the bisimilarity
as in the cycle example above.
Listing some typical valid equations for the bisimulation 
can be a shortcut \cite{Buneman,LOPSTR},
but it was only sound and \Hi{not complete} for bisimulation.

Hence, we give an algebraic and type-theoretic formulation
of UnCAL by giving equational axioms of UnCAL graphs.
In this paper, we prove completeness of 
our proposed axioms
using iteration algebra \cite{iter-alg}.
Thus we have a \Hi{complete} syntactic axiomatisations 
of the equality on UnQL/UnCAL graphs,
as a set of axioms capturing the original bisimulation,
without touching graphs, \epsilon-edges, and
the notion of bisimulation explicitly.
We prove it 
by connecting it with the algebraic axiomatisations of bisimulation
\cite{BE,Esik00}.

\subsec{How to model UnCAL and structural recursion} 
The first idea to understand UnCAL is 
to interpret it as a categorical structure.
We can regard edges as \Hi{morphisms} (of the opposite directions), the vertical composition $\at$
as the \Hi{composition  of arrows}, and \cyc as 
a \Hi{fixpoint operator} in a suitable category.
Thus the target categorical structure should have a notion of 
fixpoint,
which has been studied in iteration theories of Bloom and \Esik \cite{BE}.
In particular,
iteration categories \cite{IterCat} 
are suitable, which are traced cartesian categories \cite{JSV}
(monoidal version is used in Hasegawa's modelling of cyclic sharing theories \cite{HasseiTLCA,Hassei})
additionally satisfying the commutative identities axiom \cite{BE}
(see also \cite{alex-plot} Section 2 for a useful overview around 
this).

We also need to model UnCAL's computational mechanism: 
``structural recursion on graphs''.
The general form of the definition of structural recursive function is
\begin{Verbatim}[commandchars=\\\{\},codes=\mathcom]
                          sfun $F$($\,\ell\,$:$\,t\,$) = $e$ \x{17em}$(\blackdiamond)$
\end{Verbatim}
where $e$ can involve $F(t)$.
The graph algorithm in \cite{Buneman} 
provide a transformation of graphs that produces some computed graphs
using the definition $(\blackdiamond)$.
It becomes a function $F$ satisfying the equations 
(\cite{Buneman} Prop. 3):
\begin{equation*}
\arraycolsep = .4mm
\renewcommand{\arraystretch}{1}
\x{3em}
\begin{array}[h]{lcllllll}
  F(& {{\var y_i}} &) &=& {\var y_i}\\
  F(& { \EMP } &) &=&   \EMP  \\
  F(& { \nil } &) &=&   \nil \\
\end{array}
\qquad
\begin{array}[h]{lcllllll}
  F(& { \xsub{x}{t} } &) &=& \xsub{x}{F({t})}\\
  F(& { s \;\uniSym\; t } &) &=& F(s) \;\uniSym\; F(t)\\
  F(& \pa{  s \opl t } &) &=& \pa{F({s}) \opl F({t}) }
\end{array}
\qquad
\begin{array}[h]{lcllllllll}
F(& \quad\; \ell \co{t} \quad &)  &=&  e
\\
  F(& { s \at t } &) &=&   F({s}) \at F({t}) &\ccc(\neqmark)  &\x{2.5em}(1)\\
  F(& { \cy{t} } &) &=&   \cy{F({t})}           \quad&\ccc(\neqmark) 
\end{array}
\label{eq:my-sfun}
\end{equation*}
when $e$ does not depend%
\footnote{Here ``$e$ depends on $t$'' means that
$e$ contains $t$ other than the form $F(t)$.}
on $t$. 
This is understandable naturally as the example \code{f2} recurses structurally
the term $\mathbf{t}_G$. Combining the above categorical viewpoint, 
$F$ can be understood as a functor that preserves \cyc 
and products (thus a traced cartesian functor).
A categorical semantics of UnCAL can be given along this idea, which will
be reported elsewhere.
This idea works for simple cases of structural recursion such as \code{f2}. 

However, 
there is a {critical mismatch} between the above
categorical view and 
UnCAL's {structural recursion} of more involved cases.
Buneman et al. mentioned
a condition that the above nine equations
hold only when $e$ \Hi{does not} depend
on $t$  in (\blackdiamond). 
Two equations marked ($\neqmark$) {do not hold} in general
if $e$ \Hi{does} depend on $t$ (other seven equations do hold). 
Crucially, \code{f1} is already this case, where \code{T} appears 
as not of the form \code{f1(T)}.
The following another example shows why ($\neqmark$) do not hold:
the structural recursive function \fnaa tests whether the argument contains 
``\code{a:a:}''. 

\begin{Verbatim}[commandchars=\\\{\},codes=\mathcom,fontsize=\small]
sfun a?(L:T)  = if L=a then true:\vnil else \vnil
sfun aa?(L:T) = if L=a then a?(T) else aa?(T)
\end{Verbatim}

The definition of \code{aa?} \Hi{does} depend on \code{T} at the ``then''-clause.
Then we have the inequalities:
\begin{Verbatim}[commandchars=\\\{\},codes=\mathcom,fontsize=\small]
aa?( (a:&)$\at$(a:\{\})) = aa?( a:a:\{\} ) = true:\{\} $\,\not=\;$ \vnil = \vnil$\at$\vnil = aa?(a:&) $\at$ aa?(a:\{\})
aa?( cycle(a:&) ) = aa?( a:a:cycle(a:&) ) = true:\{\} $\,\not=\;$ \{\} = cycle(\{\}) = cycle(aa?(a:&)) 
\end{Verbatim}
This means that $F$ does not preserve \cyc in general, and even \Hi{is not functorial}, thus
the categorical view seems not helpful to understand this pattern of recursion.

\bigskip

In this paper, we consider 
\Hi{algebraic semantics} of UnCAL using the notion of iteration
\Sig-algebras \cite{iter-alg,Esik00} in \Sec \ref{sec:alg}. 
It solve the problem mentioned above, i.e.
we derive the structural recursion even when the case that $e$ depends on $t$
within the algebraic semantics.

\medskip
\subsec{Organisation}
This paper is organised as follows. We first 
give a framework of equational theory for UnCAL graphs
by reformulating UnCAL graph data in a type theoretic manner
in Section \ref{sec:theory}.
We then give algebraic semantics of UnCAL using iteration \Sig-algebras
in Section \ref{sec:alg}.
We prove completeness of our axioms for UnCAL graphs for bisimulation
in Section \ref{sec:bisim}.
We further derive structural recursion on UnCAL graphs
in Section \ref{sec:prim}.
Finally, in Section \ref{sec:bex}.
we show several examples how structural recursive functions on graphs
are modeled.


%% file: theory.tex
\newcommand{\diaNote}{%
\footnote{Identities and composite morphisms are omitted.
All directions of original edges looks
reversed in a diagram of a category.
These are just conventions of drawing a diagram.}
}

\section{UnCAL and its Equational Theory}\label{sec:theory}

We give a framework of equational theory for UnCAL graphs.
We 
reformulate UnCAL graph data in a type theoretic 
manner.
We do not employ the graph theoretic and operational concepts
(such as \epsilon-edges, bisimulation, and the graph theoretic definitions in
Fig. \ref{fig:graph-constr}).
Instead, we give an algebraic axiomatisation of  UnCAL graphs
following the tradition of categorical type theory \cite{Crole}.
The syntax in this paper
is slightly modified from the original presentation \cite{Buneman} 
to reflect the categorical idea, 
which may be more readable for the reader familiar with 
categorical type theory.

\subsection{Syntax}\label{sec:syntax}

\subsec{Markers and contexts}
We assume an infinite set of symbols called \Hi{markers}, denoted by typically
$x,y,z,\ooo$. One can understand markers as variables in a type theory.
The marker denoted by $\dmark$ is called the {default marker},
which is just a default choice of a marker having
no special property.
Let 
$L$ be a set of \W{labels}. 
A \W{label} $\ell$ is a symbol (e.g. $\code a,\code b,\code c,\ooo$ 
in Fig. \ref{fig:G}).
A \W{context}, denoted by $\seq{\v{x}_1,\v{x}_2,\ooo}$,
is a sequence of pairwise distinct markers .
We typically use $X,Y,Z,\ooo$ for contexts.
We use $\seq{}$ for the empty contexts, 
$X\conc Y$ for the concatenation, and $|X|$ for its length.
We may use the vector notation $\;\vec{\v x}\;$ for sequence $\v{x_1},\ooo,\v{x_n}$.
The outermost bracket $\seq{\;}$ of a context may be omitted.
We may use the abbreviations for the empty context $0=\seq{}$. 
Note that the concatenation may need suitable renaming to 
satisfy pairwise distinctness of markers.

\subsec{Raw terms}
\[
\begin{array}[h]{lclllllllllllllllllll}
t &::=& {\v{y}}_{\,Y} 
  &\;|\;&   \ell\c{t} 
  &\;|\;&   s \at t 
  &\;|\;&   \pa{s\opl t  }
  &\;|\;&   \cyX{t} 
  &\;|\;&   \nil_Y
  &\;|\;&   \EMP_Y &\;|\;& \mult
  &\;|\;&   \xsub{{x}}{t}
\end{array}
\]
We assume several conventions to simplify the presentation of theory.
We often omit subscripts or superscripts such as $Y$ when they are unimportant or inferable.
We identify $\pa{\pa{s \oppl t} \oppl u}$ with $\pa{s \oppl \pa{t \oppl u}}$;
thus we will freely omit parentheses as $\pa{t_1\opl \ooo \opl t_n}$.
A constant $\mult$ express a branch in a tree,
and we call the symbol $\mult$ a \W{man}, because it is similar to the shape of a kanji 
or Chinese character meaning a man,
which is originated from the figure of a man having two legs (and the
top is a head).

\subsec{Abbreviations}
We use the following abbreviations. 
\[
\begin{array}[t]{cllllllll}
\Uni{s}{t} &\deq& \mult \at \pa{s\opl t}\\
\pi_1 &\deq& \v{x}_{\seq{\vv x,\vv y}}\\
\pi_2 &\deq& \v{y}_{\seq{\vv x,\vv y}}
\end{array}
\quad
\begin{array}[t]{cllllllll}
s \X t &\deq& \pa{s\at \pi_1 \opl t\at \pi_2}\\
\Id_{\seq{\vv x}} &\deq& \v{x}_{\seq{\vv x}}\\
\Id_{\seq{x_1,\ooo,x_n}} &\deq& \v{x_{1\; \seq{\vv{x_1}}}}\X\ccc\X \v{x_{n\; \seq{\vv{x_n}}}}\\
\end{array}
\quad
\begin{array}[t]{cllllllll}
\dup_X &\deq& \pa{\Id_X \opl \Id_X}\\
\fn{c} &\deq& \pa{\pi_2\opl\pi_1}
\end{array}
\]
Inheriting the convention of $\pa{-,-}$, 
we also identify 
$(s \X t) \X u$ with $s \X (t \X u)$,
thus we omit parentheses as $t_1\X \ooo \X t_n$.

\y{-1em}
\subsection{Typed syntax}

For contexts $X,Y$,
we inductively define 
a judgment relation $\ju{t}{Y}{X}$ of terms
by the typing rules
in Fig. \ref{fig:typing}.
We call a marker \Hi{free} in $t$
when it occurs in $t$ other than the left hand-side of a definition 
$\xsub{x}{s}$.
In a judgment, free markers in $t$ are always taken from $Y$.
Thus $Y$ is a variable context (which we call the \W{source context})
in ordinary type theory,
and $X$ is the roots (which we call the \W{target context} or \W{type}).
For example, the term $\mathbf{t_G}$ in \Sec \ref{sec:review} is well-typed
\quad
$
\ju{\mathbf{t_G}} {\v{y_1},\v{y_2}}{\dmark},
$
\quad
which corresponds
a graph in Fig. \ref{fig:G},
where the marker \dmark is the name of the root.
When $t$ is well-typed by the typing rules, 
we call $t$ a (well-typed UnCAL) term.
We identify $t$ of type $\dmark$
with $\xsub{\dmark}{t}$.  

\begin{rulefigw}\small
\begin{meq}
\ninfrule{Nil}{}{
\ju{ \nil_Y  }{Y}{\dmark}
}
\qquad
\ninfrule{Emp}{
}{
\ju{ \EMP_Y  }{Y}{\seq{}}
}
\qquad
\ninfrule{Man}{
}{
 \ju{ \mult_{\seq{\var{y_1},\var{y_2}}} }{{\var{y_1},\var{y_2}}}{\dmark}
}
\qquad
\\[1em]
\ninfrule{Com}{
 \ju{ s }{Y}{Z}\qquad  \ju{ t }{X}{Y}
}{
 \ju{ s \at t }{X}{Z}
}
\qquad
\ninfrule{Label}{
 \ell \in L \infspc \ju{ t }{Y}{\dmark}  
}{
 \ju{ \ell \co{t} }{Y}{\dmark}
}
\qquad
\ninfrule{Mark}{
Y = \seq{\var{y_1},\ooo,\var{y_n}}
}{
 \ju{ {\var y_i}_Y }{Y}{\dmark}
}
\\[1em]
\ninfrule{Pair}{
 \ju{ s }{Y}{X_1}\qquad \ju{ t }{Y}{X_2}
}{
 \ju{ \pa{s \opl t} }{Y}{X_1\conc X_2}
}
\qquad
\ninfrule{Cyc}{
 \ju{ t }{Y \conc X}{ X}
}{
 \ju{ \cyX{t} }{Y}{X}
}
\qquad
\ninfrule{Def}{
\quad      \ju{ t }{Y}{ \dmark}
}{
 \ju{ \xsub{x}{t} }{Y}{ {\var x}}
}
\end{meq}

\caption{Typing rules}
\label{fig:typing}
\end{rulefigw}

\DefTitled[def:subst]{Substitution}
Let $Y=\seq{\v y_1\,\ccc,\v y_k},\, W$ be contexts such that 
$|Y| \le |W|$ and $Y$ can be embedded into $W$ in an order-preserving manner,
and $Y'$ is the subsequence of $W$ deleting all of $Y$
 (NB. $|W| = |Y| + |Y'|$, $Y'$ is possibly empty).
Suppose $\ju t W X,\qquad \ju {s_i} Z {\seq{y_i}} \qquad (1\le i \le k).$
Then a substitution $\;\ju {t \sub{\vec{{y}}}{\vec s}} {Z \conc Y'} X\;$
is inductively defined as follows.
\[
\arraycolsep = 1mm
\begin{array}[h]{lllll} 
\begin{array}[h]{rclllllllllllllllllll}
\var{y_i}        \qesub &\deq& s_i \\
\var x\;         \qesub &\deq& x \;\,\text{(if $x$ in $Y'$)}\\
\nil_Y      \qesub &\deq& \nil_{Z + Y'}\\
\EMP_Y      \qesub &\deq& \EMP_{Z + Y'}\\
(\ell\co t)    \qesub &\deq& \ell\co{(\, t\qesub\,) }
\end{array}
\;
\begin{array}[h]{rclllllllllllllllllll}
(t_1\at t_2)  \qesub &\deq& t_1\at (t_2 \qesub)\\
\pa{t_1\opl t_2} \qesub &\deq& \pa{(t_1\qesub) \opl (t_2\qesub)}\\
\cy{t}      \qesub &\deq& \cy {t \qesub}\\
\xsub{x}{t} \qesub &\deq& \xsub{x}{ t \qesub }\\
\
\end{array}
\\
\mult_{\seq{\var{y_1},\var{y_2}}} \subst{\var{y_1}\maps s_1, \var{y_2}\maps \var{s_2}} \deq
 \mult_{\seq{\var{y_1},\var{y_2}}} \at (s_1\opl s_2)
\end{array}
\]
\oDef
Note that ${t \sub{\vec{{y}}}{\vec s}}$ denotes 
a meta-level substitution operation,
not an explicit substitution.

\subsection{Equational theory}
\y{-.5em}
For terms $\ju{ s }{Y}{X}$ and $\ju{ t }{Y}{X}$, an \W{(UnCAL) equation} is of the form
$
Y \pr s = t \;\;: X.
$
Hereafter, for simplicity,
we often omit the source $X$ and target $Y$ contexts, and 
simply write $s = t$ for an equation, 
but even such an abbreviated form,
we assume that
it has implicitly suitable source and target contexts
and is of the above judgemental form.

Fig. \ref{fig:axioms} shows 
\Hi{our proposed axioms} \AxG to characterise UnCAL graphs.
These axioms are 
chosen to soundly and completely represent the original 
bisimulation of graphs by the equality of this logic.
Actually, it is sound: 
for every axiom $s = t$, $s$ and $t$ are bisimilar.
But completeness is not clear only from the axioms.
We will show it in \Sec \ref{sec:alg}.

\begin{rulefigw}
\[
\normalsize
\arraycolsep = 0mm
\begin{array}[h]{lrllllllll}
\axtit{Composition}\\
\urule{(sub1)}&    t \,\at (\nxsub{y}{s}) &\;=&\; t \sub{y}{s}\\
&&&\quad\urule{for}\;\,\;\ju{ t }{ y}{X}
\\
\axtit{Parameterised fixpoint} 
\\
\urule{(fix)} &\cy t &\;=&\; t \at \cpair{\Id_Y}{\cy t}\\
\urule{(Beki\u{c})\;} & \cy{\ccpair{t}{s}} 
&\;=&\; \cpair{\pi_2}{\cy{s}} \,\at\, \\
&\multicolumn{4}{c}{\quad\cpair{\Id_Y} {\cy{t \at \cpair{\Id_{Y\!,\! X}}{\cy{s}}}}}
\\
\urule{(nat$_Y$)} &\cy t \at s  &\;=&\;  \cy{t \at (s\X \Id_X)} \\
\urule{(nat$_X$)} &\cy{s \at t} &\;=&\;  s \at \cy{ t \at (\Id_Y \X s) } \\
\urule{(CI)} & 
\multicolumn{3}{l}{\cy{ \pa{t \at (\Id_X \X \rho_1)\opl\ooo \opl t \at (\Id_X \X \rho_m) }}}
\\
&\multicolumn{3}{l}{\qquad =\;
\Delta_m \at \cy{t \at (\Id_X \X \Delta_m)}}
\\
\end{array}
\qquad\qquad
\begin{array}[h]{lrllllllll}
\axtit{Deleting trivial cycle}
\\
\orule{(c2)} & \cy{\mult} &\;=&\; \Id \\

\axtit{Commutative monoid}
\\
\urule{(unitL$\mult$)} &\mult \at (\nil_0\X\Id) &\;=&\; \Id \\
\urule{(assoc$\mult$)} &\mult \at (\Id\X \mult)&\;=&\;\mult \at (\mult\X \Id) \\
\orule{(com$\mult$)} &\mult \at \fn{c} &\;=&\;\mult  
\\
\axtit{Degenerated bialgebra}
\\
\urule{(compa)} &\multicolumn{3}{c}{\dup\at\mult = (\mult\X\mult) \at (\Id\X\fn{c}\X\Id) \at(\dup\X\dup)} \\
\orule{(degen)} &\mult\at\dup &\;=&\; \Id
\end{array}
\]
\caption{Axioms \AxG for UnCAL graphs}
\label{fig:axioms}
\end{rulefigw}

The axiom (sub1) 
is similar to the \beta-reduction in the \lmd-calculus,
which induces the axioms for cartesian product (cf. the \textbf{derived theory} below).
The cartesian structure provides a canonical commutative comonoid
with comultiplication \Delta.

Two terms are paired with a common root by
$\Uni{s}{t}=\mult\at(s \opl t)$.
The commutative monoid axioms states that this pairing $\Uni{-}{-}$ can be 
parentheses free in nested case.
The degenerate bialgebra axioms state the 
compatibility between the commutative monoid and comonoid structures.
The degenerated bialgebra is suitable to model directed acyclic graphs
(cf. \cite{FiorePROP} \Sec 4.5), where it is stated within a PROP \cite{MacLane}.
The monoid multiplication $\mult$ expresses a branch in a tree,
while the comultiplication $\Delta$ expresses a sharing.
Commutativity expresses that there is no order between the branches of a node,
cf. (commu$\uniSym$) in the \textbf{derived theory} below,
and
degeneration expresses that the branches of a node form a set (not a sequence), 
cf. (degen').

Parameterised fixpoint axioms axiomatise a fixpoint operator.
They (minus (CI)) are known as the axioms for
Conway operators of Bloom and \Esik \cite{BE},
which ensures that 
all equalities that holds in cpo semantics
do hold.
It is also arisen in work independently 
of Hyland and Hasegawa \cite{Hassei}, who established a connection
with the notion of traced cartesian categories \cite{JSV}.
There are equalities that Conway operators do not satisfy,
e.g. $\cy{t} = \cy{t \at t}$ does not hold only by the Conway axioms.
The axiom (CI) fills this gap, which
corresponds to
the commutative identities of Bloom and \Esik \cite{BE}.
This form is taken from
\cite{alex-plot} and adopted to the UnCAL setting,
where
$\Delta_m\deq \pa{\Id_\dmark\opl\ccc\opl \Id_\dmark}$,
$Y=\seq{\var{y_1},\ooo,\var{y_m}},\; 
\ju{\Delta_m}{\dmark}{Y},\quad
\ju{t}{X \!+\! Y}{\dmark},\quad
\ju{\rho_i}{Y}{Y}$
such that $\rho_i=\pa{q_{i1}\opl\ooo\opl q_{im}}$ 
where each $q_{ij}$ is one of $\ju{\pi_i}{Y}{\dmark}$ 
for $i=1,\ooo,m$.
The axiom (c2) (and derived (c1) below) have been taken as
necessary ones for completeness for bisimulation used in several
axiomatisations, e.g. \cite{MilnerRegular,BET,Esik00}.

The equational logic \ELu for UnCAL is a logic
to deduce formally proved 
equations, called \W{(UnCAL) theorems}.
The equational logic is almost the same as ordinary one for algebraic terms.
The inference rule of the logic 
consists of reflexivity, symmetricity, transitivity, 
congruence rules for all constructors,
with the following axiom and the substitution rules.
{\small
\[
\ninfrule{Ax}{
 (\ju{ {s = t} }{{Y}}{ X}) \in E
}{
 \ju{ {s = t} }{{Y}}{ X} 
}
\qquad
\ninfrule{Sub}
{\ju {t=t'} W X\qquad \ju {s_i=s'_i} Z {\var y_i} \; (1\le i \le k)}
{\ju {t \sub{\vec{{y}}}{\vec s} = t' \sub{\vec{{y}}}{\vec s'}} {Z + Y'} X}
\]
}%
The set of all theorems deduced from the axioms \AxG is
called a \W{(UnCAL) theory}.

\newpage
\subsec{Derived theory} 
The following are formally derivable from the axioms, thus are theorems.\y{-.5em}
\begin{meqa}
&\begin{array}[h]{lcllllllll}
\text{(tmnl)}&%
\multicolumn{3}{c}{t \;=\; \EMP_Y\quad\text{for all }\ju{t}{Y}{\seq{}}}\\
\text{(fst)}& \pi_1 \at \pa{s \opl t} &\;=&\; s \\
\text{(snd)}& \pi_2 \at \pa{s \opl t} &\;=&\; t \\
\end{array}
\quad
\begin{array}[h]{lcllllllll}
\text{(dpair)} & \cpair{t_1}{t_2}\at s &\;=&\; \ccpair{t_1\at s}{t_2\at s}\\
\text{(fsi)} & \ccpair{\pi_1}{\pi_2} &\;=&\; \Id\\
\text{(SP)} & \pa{\pi_1 \at t \opl  \pi_2 \at t} &=& t\\ 
\end{array}
\\
&\begin{array}[h]{lcllllllll}
\text{(bmul)}  &\EMP_{\dmark}\X\EMP_{\dmark} &=&\EMP_{\dmark}\at \mult \\
\text{(unitR$\mult$)} &\mult \at (\Id\X\nil_0) &=& \Id \\
\text{(c1)} & \cy{\id} &=& \nil_0 \\
\text{(unR$\at$)}& t\at \Id &=& t \\
\text{(unL$\at$)}& \Id \at t  &\;=&\; t \\
\text{(assoc$\at$)}&(s \at t) \at u &\;=&\; s \at (t \at u) 
\end{array}
\begin{array}[h]{lcllllllll}
\text{(bcomul)} &\dup \at \nil_0  &\;=&\; (\nil_0\X\nil_0)\\
\text{(bunit)} &\EMP_{\dmark} \at \nil_0 &=&\Id\\
\text{(comm$\uniSym$)}& \Uni{s}{t} &=&  \Uni{t}{s} &&\\
\text{(unit$\uniSym$)}& \Uni{\nil}{t} &=& t \;=\; \Uni{t}{\nil} \\
\text{(assoc$\uniSym$)}& \Uni{\Uni{s}{t}}{u} &=& \Uni{s}{\Uni{t}{u}} \\
\text{(degen')}& \Uni{t}{t} &=&  t &&\\
\end{array}
\end{meqa}
Because of the first three lines, UnCAL has the cartesian products.
For (c1), the proof is
\[
\cy{\Id} =^{\text{(unitL}\mult)} \cy{\mult\at (\nil_0\X\Id)} =^{\text{(nat}_Y)}
 \cy{\mult}\at\nil_0 =^{\text{(c2)}} \Id\at\nil_0.
= \nil_0.
\]

\Lemma\label{th:sub}
Under the assumption of Def. \ref{def:subst},
the following is an UnCAL theorem.
\[
\arraycolsep = 1mm
\renewcommand{\arraystretch}{1.3}
\begin{array}[h]{lrllllllll}
\text{\rm (sub)}\x{5em}& t \at \cpair{s_1\opl\ccc\opl s_k}{\Id_{Y'}} = t\qesub
\x{8em}
\end{array}
\]
\oLemma


%% file: pr.tex

%% file: alg.tex
\section{Algebraic Semantics of UnCAL}\label{sec:alg}

In this section, we consider algebraic semantics of UnCAL.
We also give a complete characterisation of the structural recursion, 
where $e$ can depend on $t$ in $(\blackdiamond)$.

\subsection{Iteration \Sig-Algebras}\label{sec:iter-alg}

We first review the notion of iteration \Sig-algebras and 
various characterisation results by Bloom and \Esik.
Let \Sig be a signature, i.e.
a set of function symbols equipped with arities.
We define \mu-terms by
$$
t \;::=\; x \;\|\; f(t_1,\ooo,t_n) \;\|\; \mux t,
$$
where $x$ is a variable. We use the convention that
a function symbol $f^{(n)}\in \Sig$ denotes $n$-ary.
For a 
set $V$ of variables, we denote by $\Tm(V)$ the set of all \mu-terms
generated by $V$.
We define \AxCO 
as the set of following equational axioms:
\[
\renewcommand{\arraystretch}{1.3}
\arraycolsep = 1mm
\begin{array}[h]{lrllllllllll}
\textbf{Conway equations} &
\mux t[ s / x] &=& t[\, \mux s[t / x]\, /x \,],\\
&\mux \muy t &=& \mux t[ x / y]  \\
\axtit{Group equations associated with a group $G$}\\
&\mux (t[ 1\cdot x / x],\ooo,t[ n\cdot x / x])_1 &=& 
\muy (x[ y / x],\ooo,[ y / x])
\end{array}
\]
\noindent
Note that \W{the fixed point law}
\[
\mux t \;=\; t[\mux t /x]
\]
is an instance of the first axiom of Conway equations by taking $s = x$.
The group equations \cite{Esik-group} known as an alternative 
form of the commutative identities,
are an axiom schema parameterised by 
a finite group $(G,\cdot)$ of order $n$, whose elements
are natural numbers from 1 to $n$. 
We also note that the \mu-notation is here extended on
vectors $(t_1,\ooo,t_n)$, and $(-)_1$ denotes the first component of a vector.
Given a vector
$x=(x_1,\ooo,x_n)$ of distinct variables, the notation
$i\cdot x=(x_{i\cdot 1},\ooo,x_{i\cdot n})$ is used.

\y{-.5em}
\DefTitled[def:iter-alg]{\cite{iter-alg}}
A \W{pre-iteration \Sig-algebra} $(A,\denA{-})$ consists of an nonempty set $A$ and 
an interpretation function $\denA{-}^{(-)}:\Tm(V) \X A^V \to A$ satisfying
\begin{enumerate}
\item $\denA{x}^\rho = \rho(x)$ for each $x\in V$
\x{4em} (iii) $\denA{t}=\denA{t'} \Longrightarrow \denA{\mux t} = \denA{\mux t'}$.
\item $\denA{\,t[t_1/x_1,\ccc,t_n/x_n]\,}^\rho = \denA{t}^{\rho'}$ with
  $\rho'(x_i)=\denA{t_i}^\rho,\; \rho'(x)=\rho(x) \;\; \text{for }x\not =x_i$ 
\end{enumerate}
\oDef

A pre-iteration \Sig-algebra can be seen as a \Sig-algebra 
$(A,\set{f_A\|f\in\Sig})$ with
extra operations $\denA{\mux t}$ for all $t$.
A pre-iteration \Sig-algebra $A$ \Hi{satisfies} an equation  $s=t$ over \mu-terms,
if $\denA{s}=\denA{t}$.
Let $E$ be a set of equations over \mu-terms. 
An \W{iteration \Sig-algebra} is a pre-iteration \Sig-algebra that satisfies
all equations in \AxCO.
An \W{iteration $(\Sig,E)$-algebra} is an iteration \Sig-algebra that
satisfies all equations in $E$.
A homomorphism of iteration \Sig-algebras $h:A \to B$ is a function
such that $h \o \denA{t} = \iden{t} \o h^V$ for all $t$.
Since the variety of iteration \Sig-algebras is exactly the variety
of all continuous \Sig-algebras (\cite{iter-alg}
Introduction), 
the interpretation of ${\mux t}$ in an iteration \Sig-algebra
can be determined through it.

We now regard each label $\ell\in L$ as an unary function symbol.
Then
we consider an iteration $L\union\set{0^{(0)},+^{(2)}}$-algebra.
We define the axiom set \AxBR by
\[
\arraycolsep = .5mm
\begin{array}[h]{cclcclclllllllllllllll}
s+(t+u) &=& (s+t)+u \qquad & s+t&=&t+s \qquad & t+0=t &\\
\mux x &=& 0 & \mux (x + y) &=&
y &\text{for }y\text{ not containing }x
\end{array}
\]
and $\AxCBR \deq \AxCO\union\AxBR$.
We write
$\;\AxCBR \pr_\mu s = t\;$ if an equation $s=t$
is derivable
from \AxCBR
by the standard equational logic \ELmu for \mu-terms. 
For example, idempotency is derivable:
\[
\AxCBR \;\;\pr_\mu\;\; t + t \,=\, t
\]
The proof is $t = \mu x.(x+t) = (\mu x.(x+t)) +t = t+t$,
which uses the last axiom in \AxBR and the fixed point law.
Since \mu-terms can be regarded as a representation of process terms 
of regular behavior as Milner shown in \cite{MilnerRegular} (or synchronization trees \cite{BE}),
the standard notion of strong bisimulation between two \mu-terms can be
defined. We write $s \bisim t$ if they are bisimilar.

\ThTitled[th:Esik]{\cite{BE,iter-alg,Esik00,Esik02}}
\begin{enumerate}
\item 
The axiom set $\AxCBR$ completely axiomatises 
the bisimulation, i.e.,
$\AxCBR \pr_\mu s = t \;\;\Longleftrightarrow\;\; s \bisim t$

\item 
The set $\Tm(V)$ of all \mu-terms
forms a free pre-iteration \Sig-algebra over $V$.

\item The set \BR of all regular $L$-labeled trees having $V$-leaves
modulo bisimulation forms
a free iteration $(L\union\set{0,+},\AxBR)$-algebra over $V$
\;(\cite{Esik00} below Lemma 2, \cite{Sew95} Thm. 2).

\end{enumerate}
\oTh
Note that \BR stands for \textbf{R}egular trees modulo
\textbf{B}isimulation, and \AxBR stands 
for the axioms for regular trees modulo bisimulation.

\subsection{Characterising UnCAL Normal Forms}\label{sec:alg-sem}
\subsecn{UnCAL normal forms}
Given an UnCAL term $t$ of type \dmark, 
we compute the \Hi{normal form} of $t$
by the following three rewrite rules 
(N.B. we do not here use the other axioms)
as a rewrite system \cite{Baader}, 
which are 
oriented equational axioms taken from the derived theory, \AxG
and abbreviations.
\[
\arraycolsep = 1mm
\normalsize
\begin{array}[h]{lrllllllll}
\text{(sub)}& 
t \at \cpair{s_1\opl\ccc\opl s_k}{\Id} &=& t\qesub
\\
\text{(Beki\u{c})} &
 \cy{\ccpair{t}{s}} 
&=& \cpair{\pi_2}{\cy{s}} \,\at\, 
    \cpair{\Id_A} {\cy{t \at \cpair{\Id_{A\X V}}{\cy{s}}}}
\\
\text{(union)} &
\mult \at (s\opl t) &=& \Uni{s}{t}
\end{array}
\]
Let $\MM$ be the set of all rewriting 
normal forms by the above rules, 
which finally erases 
all $\pa{- \opl -}$ and $\at$ in a given $t$.
Normal forms are uniquely determined
because the rewrite rules are confluent and terminating,
hence have the unique normal form property \cite{Baader}.
Then by induction on terms we have that terms in $\MM$ follow the grammar 
\[
\begin{array}[h]{lclllllllllllllllllll}
\MM \ni  t ::=& {\var{y}}
  &\;\;|\;\;&   \ell\co{t} 
  &\;\;|\;\;&   \cyX{t} 
  &\;\;|\;\;&   \nil
  &\;\;|\;\;&   \Uni{s}{t} 
  &\;\;|\;\;&   \xsub{{x}}{t}.
\end{array}
\]
Any outermost definition must be of the form $\xsub{\dmark}{t'}$ 
by the assumption that the original given $t$ is of type \dmark,
thus we identity it with $t'$.
Other definitions appear inside of $t$, as the following cases:
\begin{itemize}
\item Case
$\Uni{\xsub{{x_1}}{t_1}}{\xsub{{x_2}}{t_2}}$. We identify it 
with merely $\Uni{t_1}{t_2}$, because marker names $\var{x_1} ,\var{x_2}$ are
hidden by this construction. 
\item Case $\ju{\cyc^{\var x}\xsub{x}{t'}}{Y}{\var x}$. 
We identify it with merely $\cyc^{\dmark}(t')$, because 
these are equivalent by renaming of free maker ${\var x}$. 
\end{itemize}
The \Hi{UnCAL normal forms} $\Nm$ are obtained from $\MM$ by these identifications.
It is of the form
\y{-.5em}
\[
\begin{array}[h]{llclllllllllllllllllll}
\Nm \;\ni\;\;  &t \;::=\;& {\var{y}}
  &\;\;|\;\;&   \ell\co{t} 
  &\;\;|\;\;&   \cyX{t} 
  &\;\;|\;\;&   \nil
  &\;\;|\;\;&   \Uni{s}{t} 
\\
\Tm(V)\;\ni\;\;  &t \;::=\;&
  {y}
  &\;\;|\;\;&  \ell(t)
  &\;\;|\;\;&  \mu x_1.\ooo\mu x_n.{t} 
  &\;\;|\;\;&  0
  &\;\;|\;\;&  {s}+{t}
\end{array}
\y{-.5em}
\]
Every normal form bijectively corresponds to a \mu-term in $\Tm(V)$, i.e. 
$
\Nm \iso \Tm(V),
$
because each the above construct corresponds to the lower one,
where $X=\seq{\var{x_1},\ooo,\var{x_n}}$.
Hereafter, we may identify normal forms and \mu-terms as above.
Define the pair of signature and axioms by
$$\UnCAL \;\deq\; (L\union\set{0,+},\;\;\AxBR). $$
We regard an arbitrary \UnCAL-algebra \AA as 
an \Hi{algebraic model} of UnCAL graphs.
First, we show the existence of a free model.
Define \NE to be the quotient of \Nm by the congruence
generated by $\AxCBR$.

\Prop[th:nf]\
\begin{wrapfigure}[5]{r}{.25\linewidth}\y{-3em}
\begin{diagram}[height=2em,width=3em]
\quad  V &\rTo^{\eta}& \NE\\
    &\rdTo<{\psi} & \dTo>{\psi^\sp}\\
    &             &\AA
\end{diagram}
\end{wrapfigure}
\NE forms a free iteration
\UnCAL-algebra over $V$.
Thus for any function $\psi:V \to \AA$, there exists 
an unique \UnCAL-algebra homomorphism $\psi^\sp$ such that 
the right diagram commutes, where \eta is an embedding of variables.
\oProp

\Prop
$\NE\; \iso\BR$.
\oProp
\Proof
By Theorem \ref{th:Esik} (iii).
\QED

\input{it-bisim}

\subsection{Interpretation in Algebraic Models}

To interpret UnCAL terms and equations, 
we connect two freeness results
in Thm. \ref{th:Esik}.  
Since UnCAL normal forms $\Nm$ is isomorphic to a free pre-iteration algebra 
$\Tm(V)$, it has the universal property.
Define $\Ter_\dmark$ to be the set of all well-typed UnCAL terms of 
type $\dmark$.
\begin{wrapfigure}[12]{r}{.25\linewidth}\y{-2em}
\begin{diagram}[height=2em,width=3em]
         &             &\Ter_\dmark\\
         &             &\dTo>{\fn{nf}}\\
\quad  V &\rTo^{\eta'}& \Tm(V) \iso \Nm\\
    &\rdTo^{\eta}\rdTo(2,4)_{\psi} & \dTo>{\iden{-}^{\eta}}\\
    &       & \NE &\iso \BR \\
    &             & \dTo>{\psi^\sp}\\
   &             & \AA \\
\end{diagram}
\end{wrapfigure}
We define $\fn{nf}:\Ter_\dmark\to\Nm$ by
the function to compute the UnCAL normal form of a term.
Then for any derivable equation $\ju{s=t}{Y}{X}$ in \ELu,
we have $\AxCBR \pr \sf{nf}(s) = \sf{nf}(t)$
by Lemma \ref{th:nf-corr},
thus for all assignment $\psi : V \to \AA$, 
$$\psi^\sp\iden{\fn{nf}(s)}^\eta 
=  \psi^\sp\iden{\fn{nf}(t)}^\eta$$
where 
$\eta$ and $\eta'$ are embedding of variables.
\\
\xx{1em}Since $\NE\; \iso\BR$, 
we name the isomorphisms $\uli{(-)}: \NE\; \to \BR$ and $\ol{(-)} :\BR \to \NE$. 
We write simply a normal form $t$ to denote
a representative $[t]$ in \NE.
Thus given a normal form $t$
(which is a syntactic term, always {finite}), $\;\uli{t}\;$ is a
(possibly infinite) regular tree
by obtained by expanding cycles in $t$ using fixpoints.
Conversely,
notice that since $\uli{t}$ is a tree, there are no cycles and
the original cycles in $t$ are infinitely expanded.
Since $\Nm \iso \Tm(V)$, the functions $\uli{(-)}$ 
may also be applied 
to \mu-terms.
The iteration \UnCAL-algebra $\BR$ has operations
$
  0_\BR = \uli{\nil}, \;
  +_\BR(r,s) = \uli{\uni{\, \ol r \,}{\, \ol s \,}}, \;
  \ell_\BR(r) = \uli{\ell(\ol{r})}.
$

\subsection{Deriving structural recursion of involved case}\label{sec:prim}

Next we model UnCAL's structural recursion of graphs.
We use pairs of 
``the recursive computation'' and
the history of data structure.
This is similar to the technique of paramorphism 
\cite{para}, which is a way to represent
primitive recursion in terms of ``fold'' in functional programming.
Our universal characterisation of graphs
is the key to make this possible
by the unique homomorphism 
from the free pre-iteration 
\UnCAL-algebra \Nm 
using the above analysis.

We take a term $\ju{e_\ell(v,r)}{X}{X}$ 
involving metavariables $v$ and $r$,
where
$e_\ell(F(t),t)$
is the right-hand side $e$ of $F(\ell\c t)$
in $(\blackdiamond)$ .
For example, in case of the example \code{f1} in Introduction 
(see also Example \ref{ex:retrieve}), we take
\[
\arraycolsep = .5mm
\renewcommand{\arraystretch}{1}
\begin{array}[h]{lllllllll}
e_\ell(v,r) &\deq& \code{result}\co r, \qquad  
  &e_\ell(F(t),t) &=& \code{result}\co t \qquad
&\text{if }\ell =\code{ethnicGroup}
\\
e_\ell(v,r) &\deq& v, 
  & e_\ell(F(t),t) &=&  F(t)
&\text{if }\ell \not=\code{ethnicGroup}
\end{array}
\]

We construct a \Hi{specific} iteration \UnCAL-algebra \BRR for
$\set{e_\ell(v,r)}_{\ell\in L}$.
Let $k\deq |X|$. 
Without loss of generality, we can assume that $e_\ell(v,r)$ is of the form
$\pa{t_1\opl\ccc\opl t_k}$ where every $t_i$ is a normal form.
We define the iteration \UnCAL-algebra $\BRR=\BR^k\X\BR$
having operation
\begin{meqa}
  \ell_\BRR(v,r) &= (\, \uli{ e_\ell(v, r)},\; \uli{\ell \co \ol{r}}
\,), \qquad
0_\BRR = (\vec{\uli{\nil}},{\uli{\nil}})
\end{meqa}
and $+_\BRR$
is an obvious tuple extensions of $+_\BR$.
Here $\vec\vnil$ is the $k$-tuple of $\vnil$. 
Hereafter, we will use this convention $\vec o$ of
tuple extension of an operator $o$.\phantom{MMMMMMMMMMMMMMMMMMM}
\begin{wrapfigure}[10]{r}{.25\linewidth}
\y{-1.8em}
\begin{diagram}[height=2em,width=3em]
\quad  V &\rTo^{\eta''}& \Tm(V) \iso \Nm\\
    &\rdTo^{\eta'}\rdTo(2,4)_{\eta} & \dTo>{\denBR{-}^{\eta'}}\\
    &       & \BR \\
    &             & \dTo>{\eta^\sp}\\
   &             & \BRR \\
    &             & \dTo>{\pi_1} \\
    &             & \BR^k &\!\!\!\!\!\!\iso \NE^k
\end{diagram}
\end{wrapfigure}
\quad 
Then, two freeness results in Thm. \ref{th:Esik}
 are depicted in the right diagram, where
$
\eta(x)=(\, \uli{\var{x_1}\opl\ccc\opl\var{x_k}},\, 
             \uli{\var{x}} \,)$.
Since $\Tm(V) \iso \Nm$, the interpretation in \BRR
is described as 
\begin{meqa}
  \denBRR{\var x}^\eta &= \eta(x), \quad
  \denBRR{\nil}^\eta = 0_\BRR, 
\quad
  \denBRR{\Uni{s}{t}}^\eta = \denBRR{s} +_\BRR \denBRR{t}
\\
  \denBRR{\ell \co t}^\eta &= \ell_\BRR(\denBRR{t}^\eta),\quad
  \denBRR{\cyc (t)}^\eta = \eta^\sp( \uli{\cyc(t)})
\end{meqa}
Now $\denBRR{-}^\eta$ is characterised as
the unique pre-iteration $L\union\set{0,+}$-algebra homomorphism from $\Tm(V)$
that extends \eta.
Defining $$\phi \deq \pi_1\o\denBRR{-}^\eta: \Nm\rTo \BR^k \iso \NE^k,$$
it is the unique function satisfying
\begin{meqa}
  \phi(\var x) &= (\uli{\var{x_1}},\ooo,\uli{\var{x_k}}), \qquad \phi(\nil) = \vec{\uli{\nil}}, \qquad
  \phi(\Uni{s}{t}) = \uli{\phi(s) \;\;\vec\uniSym\;\; \phi(t)},\\
  \phi(\ell \co t) &= \uli{ e_\ell(v,t) },\qquad
  \phi(\cyc (t)) = \pi_1\o\eta^\sp( \uli{\cyc(t)})
\end{meqa}
The function \phi takes normal forms of the type \dmark. 
For non-normal forms, just precompose \fn{nf}, i.e.,
define the function $\Phi : \Ter_\dmark \to \NE^k$ 
by $\Phi(s) \deq \phi(\fn{nf}(s))$,
thus, 
  $\Phi^{|X|}:\Ter_X \to   {\NE}^{k|X|} \to \Ter_X^k$,
because $\Ter_X \iso \Ter_\dmark^{|X|}$.
In summary, we have the following, where $s$ is a possibly non-normal form
\[
\arraycolsep = .3mm
\x{1em}
\begin{array}[h]{llllllllllllllllllllllllll}
\Phi(s) &=& \phi(\, \fn{nf}(s) \,) &&
\phi(\var x) &=& \pa{\var{x_1}\opl\ccc\opl\var{x_k}}\quad
&\phi(\nil) = \vec{\nil}
\\
  \Phi^{|X|+|Y|}(\pa{t_1 \opl t_2}) &=& 
   \Phi^{|X|}(t_1) \;\vec\X\; \Phi^{|Y|}(t_2)\;\;&\quad\ &
  \phi(\ell \co t) &=& { e_\ell( \phi(t),\, t ) }
&
  \phi(t_1 \,\uniSym\, t_2) = \phi(t_1) \;\;\vec\uniSym\;\; \phi(t_2)
&\x{1em}(2)
\\
\Phi^0(\EMP) &=& \EMP  &
&
\phi(\cyc (t)) &=& \pi_1\o\eta^\sp( \uli{\cyc(t)})\quad
  &
\end{array}
\]
where $\vec\times$ 
is the ``zip'' operator of two tuples.
Here we use
a map $\NE \to \Term(V)$ to regard a normal form modulo \AxCBR
as a term,
for which any choise of representative is harmless,
because UnCAL graphs are identified by bisimulation and
\AxCBR axiomatises it.
Identifying three kinds functions $\Phi, \Phi^{|X|}, \phi$ as a 
single function (also denoted by \Phi, by abuse of notion) on $\Term(V)$, 
this \Phi is essentially {what 
Buneman et al. \cite{Buneman}
called the structural recursion on graphs for the case
that $e$ depends on $t$}.
Actually, we could make the characterisation
{more precise} than \cite{Buneman}, 
i.e., we obtain also the laws for %
the cases of $\at$  (by the case $\Phi(s) = \phi(\fn{nf}(s))$)
and \cyc, 
which tells how to compute them.

This is not merely rephrasing the known result, but also
a {stronger characterisation}, which gives 
precise understanding of the structural recursion on graphs:

\begin{enumerate}
\item Buneman et al. stated that 
(\ref{eq:my-sfun}) without $(\neqmark)$ 
is a \Hi{property} (\cite{Buneman} Prop. 3) of 
a ``structural recursive function on graphs'' defined by
the algorithms in \cite{Buneman}.
This property (i.e. soundness) is desirable, 
but unfortunately, no completeness was given.
There may be many functions that satisfy the {property}.
In contrast to it, our characterisation is sound and \W{complete}:
(2) determines a \Hi{unique} function by the universality.

\item This derivation does not entail
$\Phi(s\at t)= \Phi(s)\at \Phi(t)$.
It tells us that the only way to compute $\Phi(s\at t)$ is to 
compute the normal form of $s\at t$ and then apply \phi.

\item This analysis does not entail
$\Phi(\cyc(t))= \cyc(\Phi(t))$ either.
The iteration algebra structure tells us that
the homomorphism \phi maps
a term $\cyc(t)$ to its interpretation in \BRR where
the cycles are expanded in a regular tree and at the same time,
labels $\ell$ are interpreted using the operations of \BRR.

\item The structure preserved by structural recursion 
is the \Hi{(pre-)iteration algebra
structure}.
The structural recursive function \phi is the composition of a pre-iteration algebra homomorphism,
an a iteration algebra homomorphism and a projection.
\end{enumerate}


%% file: it-bisim.tex
\subsection{Completeness of the Axioms for Bisimulation}
\label{sec:bisim}
Buneman et al. formulated that
UnCAL graphs  were identified by 
\Hi{extended bisimulation},
which is a bisimulation on graphs involving
\epsilon-edges. 
As discussed in \Sec \ref{sec:review}, 
since our approach is to use only UnCAL terms,
it suffices to consider only the standard (strong) bisimulation  between
UnCAL terms, as done in \cite{MilnerRegular, BE, Esik00, Esik02}.
We denote by $\bisim$ bisimulation for UnCAL term.

In this subsection, we show the completeness of \AxG for bisimulation,
using the following Lemma \ref{th:nf-corr} that reduces the problem of
\ELu to that of \ELmu through UnCAL normal forms.
\AxCBR has been shown to be complete for the bisimulation \cite{BE}.

\Lemma[th:nf-corr]
For UnCAL normal forms $n,m\in\Nm\;$,
$\AxCBR \pr_\mu n = m \Longleftrightarrow\;$ 
$\ju{n=m}{Y}{X}$ is derivable from \AxG in \ELu.
\oLemma
\Proof
$[\Rightarrow]:$
By induction on proofs of \ELmu.
For every axiom in \AxCBR, there exists the corresponding axiom
in \AxG or an \ELu theorem, hence it can be emulated.

\noindent
$[\Leftarrow]:$
By induction on proofs of \ELu.
Let $s = t$ is an axiom of \ELu.
It easy to see that
taking normal forms of both side, they are 
equal term, or correspond to an axiom in \AxCBR or \ELmu theorem.
\QED

\ThTitled[th:complete]{Completeness} 
\AxG is sound and complete for
the bisimulation, i.e.,\\
$\ju{s=t}{Y}{X}\;$ is derivable from \AxG in \ELu\; iff\;
$\; s \bisim t  $.
\oTh
\Proof
$[\Rightarrow]:$
Because every axiom in \AxG is bisimilar, and 
the bisimulation is closed under contexts
and substitutions \cite{Buneman}.

\noindent
$[\Leftarrow]:$
Suppose $s \bisim t$. Since for each rewrite rule for
the normalisation function \fn{nf},
both sides of the rule is bisimilar, $\fn{nf}$ preserves the bisimilarity. 
So we have
$s \bisim \fn{nf}(s)\bisim  \fn{nf}(t) \bisim t$.
Since \AxCBR is complete axioms of bisimulation \cite{BE,Esik00}, 
$\AxCBR \pr_\mu \fn{nf}(s) =  \fn{nf}(t)$.
By Lemma \ref{th:nf-corr}, we have a theorem $\ju{\fn{nf}(s)=\fn{nf}(t)}{Y}{X}$.
Thus $s=t$ is derivable.
\QED


%% file: bex.tex
\subsection{Examples}\label{sec:bex}

We may use the notation $\{t_1,t_2,\ooo \}$ as the abbreviation of
$\Uni{t_1}{t_2} {\;\RM{\scriptsize $\union$}\;} \ccc$.

\ExampleTitled{\cite{Buneman} Replace all labels with \code{a}}
This is the example considered in Introduction.
\begin{verbatim}
  sfun f2(L:T) = a:f2(T)
\end{verbatim}
In this case, the recursion does \Hi{not} depend on \code T
(because the right-hand side uses merely \code{f2(T)}).
We define the iteration \UnCAL-algebra \BRR by
$$
\ell_\BRR(v,r) = ({\code{a}\c v} ,\; \ell \c r).
$$
(We may omit over and underlines to denote the isomorphisms for simplicity).
Then \Phi is the desired structural recursive function \code{f2}.
E.g.
\begin{meqa}
\Phi(\code b \co \cy{\code c \co \dmark}) 
\;=\; \code a\co \phi(\cy{\code c \co \dmark})
\;=\; \code a\co \pi_1 \o \eta^\sp(\code c\co\code c\co\ccc)
\;=\; \code a\co (\ol{\code a\co\code a\co\ccc})
\;=\; \code a\co \cy{\code a\co\dmark}
\end{meqa}

\oExample

\ExampleTitled{\cite{Buneman} Double the children of each node}
\begin{Verbatim}[commandchars=\\\{\},codes=\mathcom]
  sfun f4(L:T) = \Uni{a:f4(T)}{b:f4(T)}
\end{Verbatim}
Example of execution.
\begin{Verbatim}[commandchars=\\\{\},codes=\mathcom,fontsize=\small]
f4(a:b:c\nnil) 
\narone \Uni{a:\{ a:\{a\nnil, b\nnil\}, b:\{a\nnil, b\nnil\} \}}{b:\{ a:\{a\nnil, b\nnil\}, b:\{a\nnil, b\nnil\} \}}
\end{Verbatim}
This case does \Hi{not} depend on \code T.
We define the iteration \UnCAL-algebra \BRR by
$$
\ell_\BRR(v,r) =(\Uni{\code{a}\c v}{\code{b}\c v},\; \ell \c r). 
$$
Then $\Phi$ gives
the structural recursive function defined by \code{f4}.
\oExample

\ExampleTitled[ex:retrieve]{\cite{Buneman} Retrieve all ethnic groups}
We revisit the example given in \Sec \ref{sec:review}.\\
For the structural recursive recursive definition of \code{f1},
\begin{verbatim}
  sfun f1(L:T) = if L = ethnicGroup then (result:T) else f1(T)
\end{verbatim}
This case \Hi{does} depend on \code T.
Example of execution:
\begin{Verbatim}[commandchars=\\\[\],codes=\mathcom,fontsize=\small]
f1(sd) \narone {result:"Celtic"\nnil, result:"Portuguese"\nnil, result:"Italian"\nnil}
\end{Verbatim}
We define the iteration \UnCAL-algebra \BRR by
\begin{meqa}
\code{ethnicGroup}_\BRR(v,r) &
\deq (\code{result}\co r,\; \code{ethnicGroup}\co r)
\\
\ell_\BRR(v,r) &\deq (v,\; \ell\co r) \text{ for }\ell \not=\code{ethnicGroup}
\end{meqa}
Then \Phi is the structural recursive function defined by \code{f1}:
\begin{Verbatim}[commandchars=\\\[\],codes=\mathcom,fontsize=\small]
\Phi(sd) = {result:"Celtic"\nnil, result:"Portuguese"\nnil, result:"Italian"\nnil}
\end{Verbatim}
\oExample

\Example
Consider another example in \Sec \ref{sec:review} of \code{aa?}.
This case \Hi{does} depend on \code T.
We define the iteration \UnCAL-algebra \BRR by
\begin{meqa}
  \code{a}_\BRR(v,r) &\deq (\code{a?}(r),\;\code{a}\co r)\\
  {\ell}_\BRR(v,r) &\deq (v,\;\ell\co r) \quad\text{for }\ell \not=\code{a}.
\end{meqa}
Then $\Phi$ gives the structural function \code{aa?} 
\begin{meqa}
&\Phi\code{((a:\&)@(a:\{\})) = }\phi(\; \fn{nf}\code{((a:\&)@(a:\{\}))} \;) = \phi\code{(a:a:\{\}) = true:\{\}}
\\
&\Phi(\code{cycle(a:\&)}) = \pi_1\o\eta^\sp(\uli{\code{cycle(a:\&)}})
= \pi_1\o\eta^\sp({\code{a:a:}\ccc}) 
= \pi_1 \,(\code{a?}({\code{a:}\ccc}), {\code{a:}\ccc})
= \code{true}:\nil
\end{meqa}
\oExample


%% file: concl.tex
\section{Conclusion}
In this paper, we have shown an application of Bloom and \Esik's iteration algebras
to model graph data used in UnQL/UnCAL for
describing and manipulating graphs.
We have formulated UnCAL and given an
axiomatisation of UnCAL graphs that characterises the original bisimulation.
We have given algebraic semantics using 
Bloom and \Esik's iteration iteration algebras.  
The main result of this paper was to show that completeness of our
equational axioms for UnCAL for the original bisimulation of UnCAL graphs via
iteration algebras.
As a consequence, we have given a clean
characterisation of the computation mechanism of UnCAL, called
``structural recursion on graphs'' 
using free iteration algebra.


%% file: ack.tex
\subsec{Acknowledgments}
I am grateful to  Kazutaka Matsuda and Kazuyuki Asada
for discussions about UnCAL and its interpretation,
and their helpful comments on a draft of the paper.
A part of this work was done
while I was visiting National Institute of Informatics
(NII) during 2013 -- 2014.
